\documentclass[aps,prb,longbibliography,showpacs,twocolumn,superscriptaddress,amsmath,amssymb,verbatim]{revtex4-1}
\usepackage[utf8]{inputenc}
\usepackage{amsmath,amsfonts,amssymb}
\usepackage{pifont}
\usepackage{rotating}
\usepackage{graphicx}    % Include figure files
\usepackage{epstopdf}
\usepackage{color}
\usepackage[caption=false]{subfig}
\usepackage{soul}
\usepackage{ulem}
\usepackage{hyperref}
\definecolor{dark-red}{rgb}{0.9,0.15,0.15}
\definecolor{dark-blue}{rgb}{0.15,0.15,0.4}
\definecolor{medium-blue}{rgb}{0,0,0.5}
\hypersetup{
    colorlinks, linkcolor={black},
    citecolor={dark-blue}, urlcolor={medium-blue}
}
\begin{document}

\title{Magnetic transition in the $J_\textrm{eff} = 1/2$ Kagom\'{e} system   Sm$_3$Sb$_3$Zn$_2$O$_{14}$}

\author{Vinod Kumar}
\email{vinodkiitb@gmail.com}
\affiliation{Department of Physics, Indian Institute of Technology Bombay, Mumbai 400076, India}

\author{S. Kundu}
\affiliation{Department of Physics, Indian Institute of Technology Bombay, Mumbai 400076, India}

\author{M. Baenitz}
\affiliation{Max Planck Institut für Chemische Physik fester Stoffe, Nöthnitzer Strasse 40, 01187 Dresden, Germany}

\author{A. V. Mahajan}
\email{mahajan@phy.iitb.ac.in}
\affiliation{Department of Physics, Indian Institute of Technology Bombay, Mumbai 400076, India}

%%-------------------------------
%               ABSTRACT
%%-------------------------------

\begin{abstract}
 We present a study of Sm$_3$Sb$_3$Zn$_2$O$_{14}$ (SSZO) through magnetization and specific heat capacity $C_p(T)$ measurements. SSZO contains well separated planes of Sm$^{3+}$ magnetic ions on a Kagom\'{e} lattice. Though our magnetic susceptibility (where the data are limited down to 2\,K) does not show any anomaly, the $C_p(T)$ data which have been taken down to 0.35 K reveal a broad maximum at $T\sim1.5$\,K  followed by a sharp peak at $T\sim0.5$\,K indicative of short-range correlations and long-range order, respectively. 
          
\end{abstract}

\date{\today}
%\pacs{75.50.Bb, 75.50.Cc, 61.43.-j, 85.75.-d, 31.15.A}

\maketitle

\section{Introduction}  

Following the discovery of quantum spin liquid behavior in herbertsmithite [ZnCu$_3$(OH)$_6$Cl$_2$]~\cite{shores_2005_JACS,Helton2007,han_2012_Nature} systems with a Kagom\'{e} geometry with $S = 1/2$ magnetic ions have become a playground for exploring novel magnetism.  There have also been several investigations on systems with an effective Kagom\'{e} network containing magnetic rare earth ions. The pyrochlore lattice is also relevant to this discussion. The pyrochlore mineral~\cite{Gardner_2010_RMP,khomskii_Book_2014} has the general formula  A$_2$B$_2$O$_7$ and is made up of corner shared tetrahedra where A and B cations form a  staggered network of vertex sharing tetrahedra. We can also express this formula as  A$_4$B$_4$O$_{14}$.  With inequivalent ions at the A and the B sites such as in (A$_3$A$^{\prime}$)(B$_3$B$^{\prime}$)O$_{14}$ and, in addition, if we replace A$^{\prime}$ and B$^{\prime}$ with non-magnetic ions then the structure contains alternating Kagom\'{e} layers,~\cite{grey_JSSC_2003} of A and B ions. With this general motivation, systems such as  RE$_3$Sb$_3$Co$_2$O$_{14}$ (RE$=$ La, Pr, Nd, Sm-Ho) have been investigated~\cite{Li_2014_JSSC,fu_JSSC_2014}. The above chemical formula can be written as [RE$_3$Co][Sb$_3$Co]O$_{14}$ where there are two distinct Kagom\'{e} lattices made of RE$^{3+}$ and Sb$^{5+}$ with Co$^{2+}$ at the center of hexagon. Replacement of the magnetic Co ions by non-magnetic  Mg and Zn is reported~\cite{Dun_PRL_2016,dun_PRB_2017}. Thus, one obtains well separated RE$^{3+}$ Kagom\'{e} magnetic planes in case  RE$^{3+}$ is magnetic.  A resulting system Tm$_3$Sb$_3$Zn$_2$O$_{14}$~\cite{ding_PRB_2018} has been reported to be a spin liquid\,\cite{balents_2010_Nature} which motivated us to take interest in this pyrochlore family with rare-earth ions forming the Kagom\'{e} layers.

Our focus here is on Sm$_3$Sb$_3$Zn$_2$O$_{14}$ (SSZO) which again can be thought of as (Sm$_3$Zn)(Sb$_3$Zn)O$_{14}$ where Sm$^{3+}$ (magnetic) and Sb$^{5+}$ (non-magnetic) form alternating Kagom\'{e} layers. Among the various lanthanides, Sm is interesting for the following reasons. The ground state of isolated Sm$^{3+}$ is $^6$H$_{\frac{5}{2}}$ due to the strong spin-orbit coupling which is active here.  In a cubic crystalline electric field, the $J=5/2$ state splits into a doubly degenerate Kramers doublet and a quartet ground state\,\cite{Abragam_Bleany,Lea_1962_JPCS}. At low temperatures, where one can ignore thermal excitations, the quartet is fully occupied and the  doublet is half filled and the physics of this system will be that of a pseudo spin-half Kagom\'{e} lattice.  The susceptibility measurements reported~\cite{sanders_JMCC_2016} on SSZO find an antiferromagnetic Curie-Weiss temperature and no anomaly down to 2 K suggesting possible spin liquid behaviour.

 We report here   the magnetic susceptibility and heat capacity  of Sm$_3$Sb$_3$Zn$_2$O$_{14}$ (SSZO). Magnetization measurements down to 2\,K show an increase in a Curie-Weiss manner without any anomaly. However, a sharp peak in the heat capacity at $T\sim0.5$\,K is indicative of long-range order. The entropy change on going from about 10 K to our lowest temperature of 0.35 K is about 50\% of R$\ln2$ and the remaining entropy is perhaps lost at lower temperatures.

\section{Sample Preparation and experimental details}
Sm$_3$Sb$_3$Zn$_2$O$_{14}$ (SSZO) has been prepared by conventional solid state reaction methods. We took initial materials, Sm$_2$O$_3$ (Alfa Aesar 99.9\%), Sb$_2$O$_3$ (Alfa Aesar 99.9\%), ZnO (Aldrich Chemicals 99.9\%) in stoichiometric amounts. We thoroughly mixed the initial compounds in a mortar-pestle and we initially fired the mixture at 900$^{\circ}$C in air. Finally we sintered the sample at 1150$^{\circ}$C in air with some intermediate grindings. After preparing the sample, to check its phase purity and to extract the crystal structure parameters, room temperature x$-$ray diffraction (XRD) measurement has been performed with a PANalytical X'PertPRO diffractometer using Cu-K$_{\alpha}$ radiation ($\lambda=1.54182$ \AA). Magnetization ($M$) measurements were performed in the temperature range 2-300 K and in various applied magnetic fields between zero and 70\,kOe, with a Quantum Design SQUID Vibrating Sample Magnetometer (SVSM). The heat capacity $C_p(T)$ measurements have been performed using the heat capacity option in a Quantum Design PPMS in the $T$-range 0.35-220\,K, in applied magnetic fields ranging from 0-90 kOe.

\section{Experimental Results}
\subsection{XRD Analysis and Crystal structure}

\begin{figure}[h!]
\centering
\includegraphics[width=8.5cm, height=6.6cm]{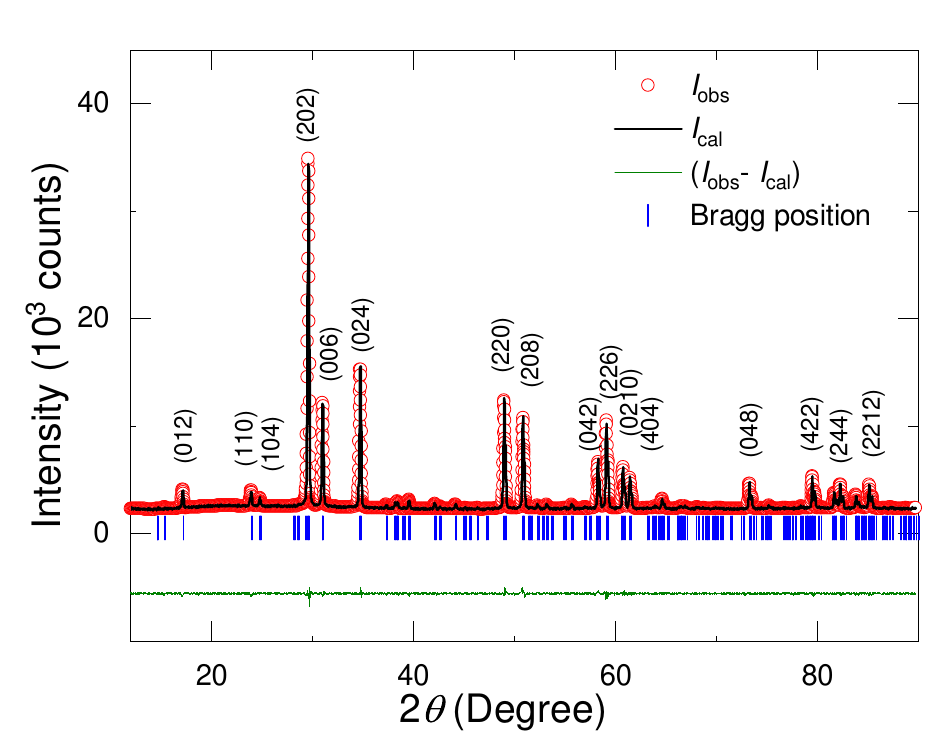}
\caption{X-ray diffraction pattern of Sm$_3$Sb$_3$Zn$_2$O$_{14}$ at 300\,K with its Rietveld refinement by FULLPROF is shown. The observed intensity is shown by open circles while its calculated intensity is the solid black line and the difference between these is shown by the solid green line at the bottom. The Bragg peak positions are shown by vertical bars along with some selected Miller indices (hkl).}
\label{fig:Fig-1}
\end{figure}

The room temperature x-ray diffraction pattern of SSZO is shown in Fig.\,\ref{fig:Fig-1} together with  Rietveld refinement using FULLPROF suite\cite{CARVAJAL_Refine}. SSZO crystallizes in a trigonal crystal symmetry with the space group $R$\,\={3}\,m (Space Group no. 166). After refinement we obtained the cell parameters $a=b=$7.43117(3)\,$\rm\AA$, $c=17.28968(11)$\,$\rm\AA$ and $\alpha=\beta=$90$^{\circ}$, $\gamma=120^{\circ}$. The extracted lattice and refinement parameters are given in Table \ref{tab:refinement parameters} and the atomic positions in Table \ref{tab:atomic_position}. These are in reasonable agreement with literature~\cite{sanders_JMCC_2016}. 

\begin{figure*}
\centering
\includegraphics[width=14cm, height=10.77cm]{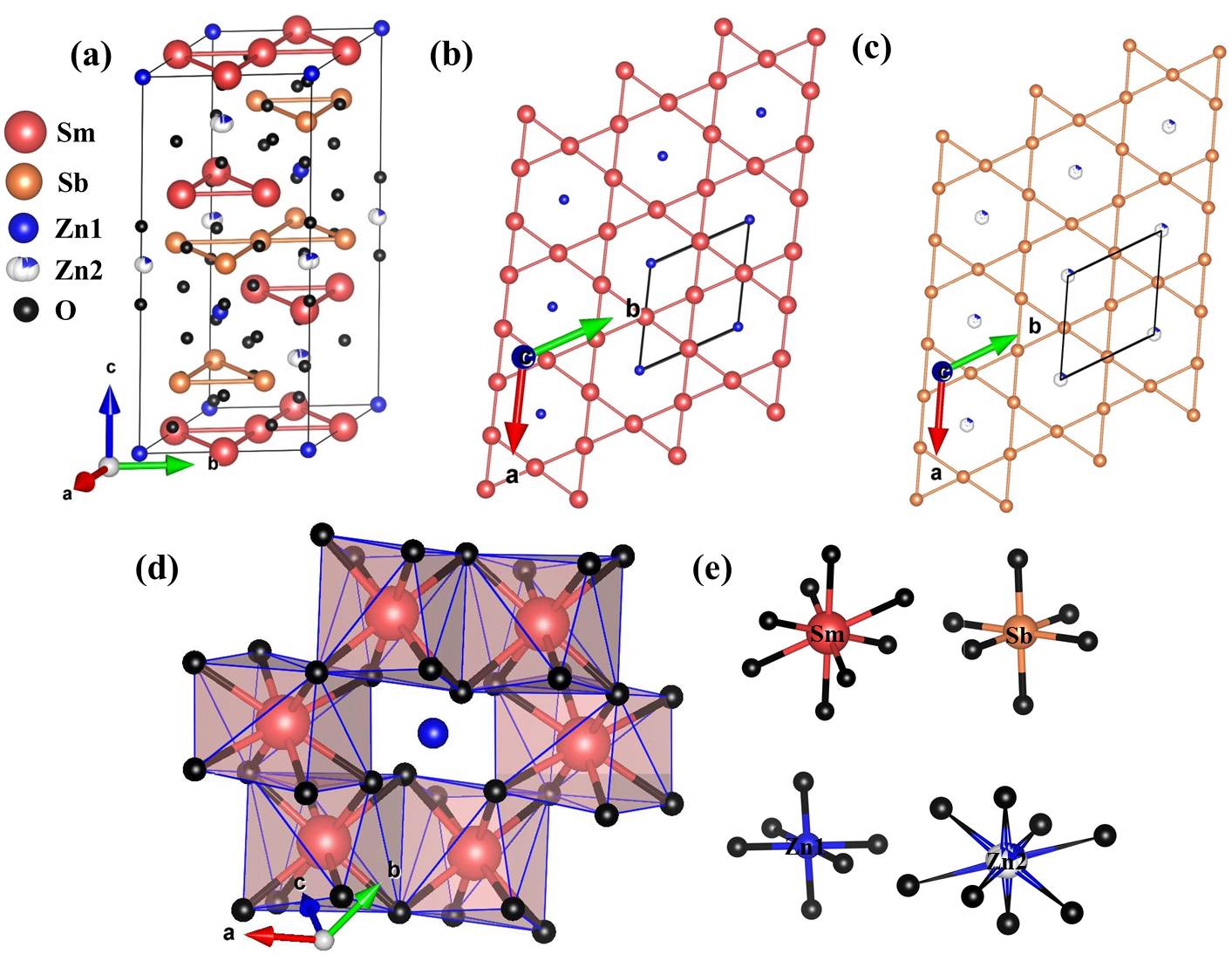}
\caption{Crystal structure of Sm$_3$Sb$_3$Zn$_2$O$_{14}$ is shown. (a) The unit cell of SSZO with Sm and Sb triangles. (b) Corner shared triangles of Sm$^{3+}$ ions forming the Kagom\'{e} structure with Zn1 at the center of hexagon. (c) Corner shared triangles of Sb$^{5+}$ ions forming a Kagom\'{e} structure with Zn2 at the center of hexagon. (d) The distorted cubic environment around Sm$^{3+}$ ions formed by oxygens, which are responsible for CEF, and Zn1 at the center of hexagon, formed by six Sm$^{3+}$ ions. (e) The local co-ordinations with O of Sm, Sb, Zn1 and Zn2 are shown.}
\label{fig:Fig-2}
\end{figure*}

 In Fig.\,\ref{fig:Fig-2} the unit cell of SSZO and various salient features are illustrated. Fig.\,\ref{fig:Fig-2}\,(d) shows the local environment around the Sm$^{3+}$ ions formed by oxygens, which is responsible for the crystalline electric field which causes splitting of the $J=5/2$ state. The eight O$^{2-}$ ions around a Sm$^{3+}$ form a distorted cube.

\begin{table}[h!]
	\centering
	\begin{tabular}{|c|c|}
		\hline Space group & $R$ \={3} m \\ 
		\hline Crystal system & Trigonal \\ 
		\hline Lattice parameters & $a$=$b$= 7.43117(3) \AA,\\& $c$=17.2897(1) \AA \\ 
		&$\alpha=\beta= 90 ^\circ, \gamma= 120 ^{\circ}$\\ 
		\hline Refinement parameters & $R_{p}=11.5, R_{wp}=7.22 $\\
		& $R_{exp}=6.7, \chi^2=1.16$\\ 
		\hline 
	\end{tabular} 
	\caption{The lattice parameters and refinement parameters after crystal structure refinement of SSZO by FULLPROF.}
	\label{tab:refinement parameters}
\end{table}

\begin{table}[h!]
	\centering
	\begin{tabular}{|c|c|c|c|c|c|c|}
		\hline Atom &Wyc. &  x & y & z & $U$(\AA$^2$)& Occ. \\ 
		\hline Sm   & 9d & 0.5000 & 0.0000 & 0.0000& 0.001(16)& 1 \\ 
		\hline Sb  & 9e & 0.5000 & 0.0000 & 0.5000 & 0.001(18) & 1 \\ 
		\hline Zn1   & 3a & 0.0000 & 0.0000 & 0.0000 & 0.007(3) & 1 \\ 
		\hline Zn2  & 18g& 0.022(1) & 0.0000 & 0.5000 & 0.005(94) & 0.16667 \\ 
		\hline O1   & 6c& 0.0000 & 0.0000 & 0.3962(9) & 0.006(1) & 1 \\ 
		\hline O2   &18h & 0.5234(6) & 0.4767(6) & 0.1441(3) & 0.004(2) & 1 \\ 
		\hline O3   &18h & 0.1380(5) & 0.8621(5) & -0.0522(4) & 0.005(2) & 1 \\ 
		\hline 
	\end{tabular} 
	\caption{The atomic positions of SSZO obtained after Rietveld refinement of the XRD data measured at 300\,K by FULLPROF.}
	\label{tab:atomic_position}
\end{table}

\subsection{Magnetization}

\begin{figure}[h!]
\centering
\includegraphics[width=9cm, height=7.2cm]{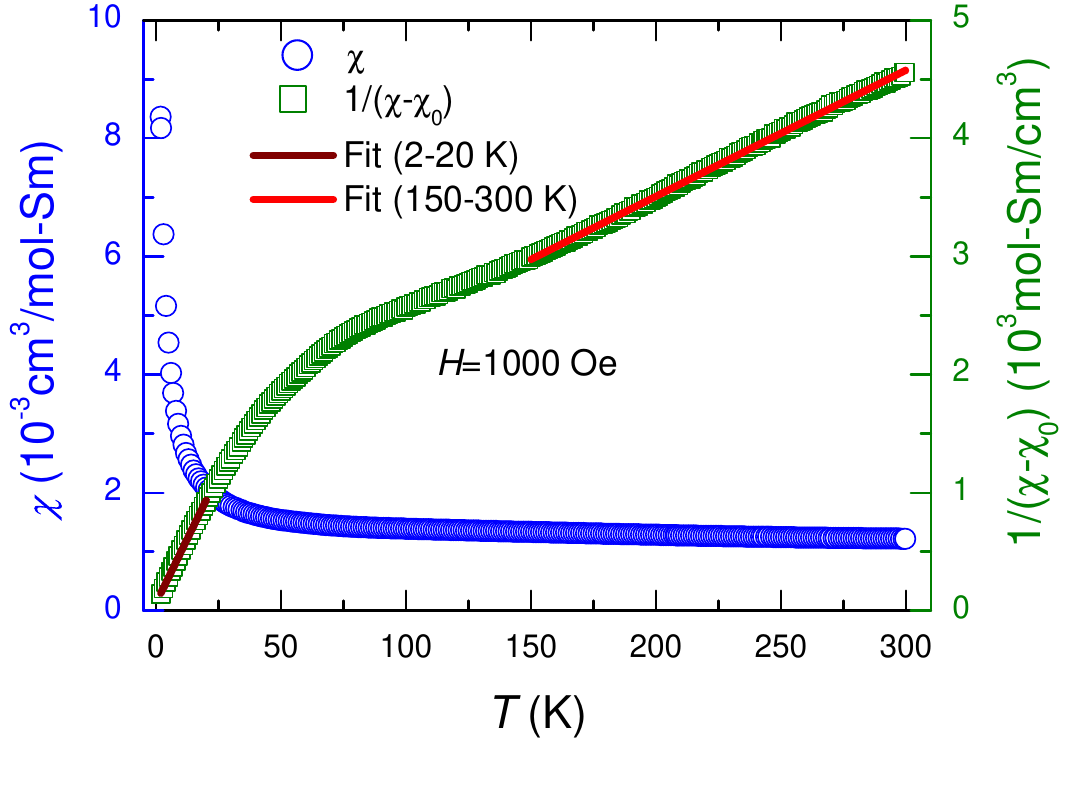}
\caption{On the left $y-$ axis $\chi(T)$ is shown as a function of $T$ at 1000 Oe for SSZO. On the right $y-$axis 1/($\chi-\chi_0$) is shown as a function of $T$. The linear fitting is done in high temperature range (150-300)\,K and low temperature range (2-20)\,K}
\label{fig:Fig-3}
\end{figure}

The magnetic susceptibility ($\chi=\frac{M}{H}$) for SSZO in 1000\,Oe  is shown as a function of temperature $T$ in Fig.\,\ref{fig:Fig-3}. The $\chi(T)$ is nearly constant from 300 to 150\,K and increases on further lowering the temperature. Fitting the $\chi(T)$ data  in the $T$-range $150-300$\,K with the Curie-Weiss (CW) law;
\begin{equation}{\label{:eq.1}}
\chi = \chi_0 + \frac{C}{(T-\theta_{CW})}
\end{equation} 
yields $\chi_0=9.94\times10^{-4}$\,cm$^3$/mol-Sm, $C$ = 0.095\,cm$^3-$K/mol-Sm, and $\theta_{CW}$ = -129 K, where $\chi_0$ is the  $T$-independent magnetic susceptibility, $C$ is the Curie constant and $\theta_{CW}$ is the Curie-Weiss temperature. The inferred effective moment is $\mu_\textrm{eff}$ =  0.87 $\mu_B$ which is close to the expected value of 0.84\,$\mu_B$ for isolated Sm$^{3+}$.  This suggests that the thermal energy is comparable to the crystal field splittings in this range and both the doublet and the quartet states are populated. Therefore to probe the magnetism associated with the ground state we have fitted the low-temperature ($2-20$\,K) region in the  $1/(\chi-\chi_0)$ vs. $T$ plot (see Fig.\,\ref{fig:Fig-3}) to a linear equation. The extracted $\theta_{CW}=-1.4$\,K  suggests  weak antiferromagnetic interactions in SSZO. Further, the above fit yields an effective magnetic moment of 0.43\,$\mu_B$. This is close to the expected value of $\mu_\textrm{eff}$ = 0.41 $\mu_B$ for the $\Gamma_7$ Kramers doublet~\cite{Meyer_2008_PRB,surjeet_singh_PRB_2008} suggesting that SSZO is a pseudo $S=1/2$ Kagom\'e system.  These results are similar to those of Ref.~\cite{sanders_JMCC_2016}.
%This shows that $J=\frac{1}{2}$ is the possible ground state Kramer's doublet in SSZO\,\cite{Lea_1962_JPCS,Meyer_2008_PRB,surjeet_singh_PRB_2008}. This analysis is also supported by magnetic entropy change data (see Fig.\,\ref{fig:dSm_0_Oe} in heat capacity section).                  

\subsection{Heat capacity}

\begin{figure}[h!]
\centering
\includegraphics[width=9cm, height=7cm]{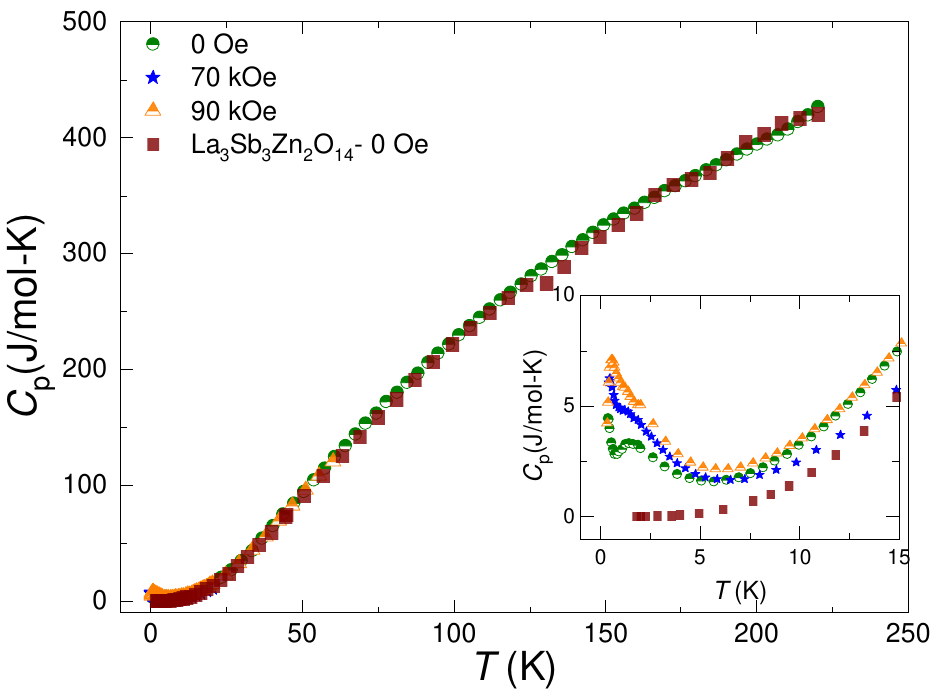}
\caption{The heat capacity $C_p(T)$ data for SSZO is shown in $T$-range 0.35$-$220 K at zero field, 70\,kOe and 90\,kOe together with the zero field $C_p(T)$ data for nonmagnetic reference La$_3$Sb$_3$Zn$_2$O$_{14}$ in $T$-range 2-220 K.  In the inset, an enlarged view of the low-temperature heat capacity data is shown.}
\label{fig:Fig-4}
\end{figure}

We have then performed heat capacity $C_p(T)$ measurements down to 0.35 K to check for signs of long-range order. We measured the heat capacity of a sintered polycrystalline pellet of mass 13.44\,mg, using the one-tau method with 1\% temperature rise. In Fig.\,\ref{fig:Fig-4} the $C_p(T)$ data of SSZO is shown in the $T$-range 0.35$-$220\,K in different applied fields together with the $C_p(T)$ data of La$_3$Sb$_3$Zn$_2$O$_{14}$ (LSZO) in the $T$-range 2$-$220\,K in zero field. LSZO is a nonmagnetic structural analog of SSZO and we will use $C_p(T)$ data of LSZO to extract the lattice heat capacity of SSZO and eventually infer the magnetic specific heat\,\cite{gopal2012specific} of SSZO. 

\begin{figure}[h!]
\centering
\includegraphics[width=9cm, height=7cm]{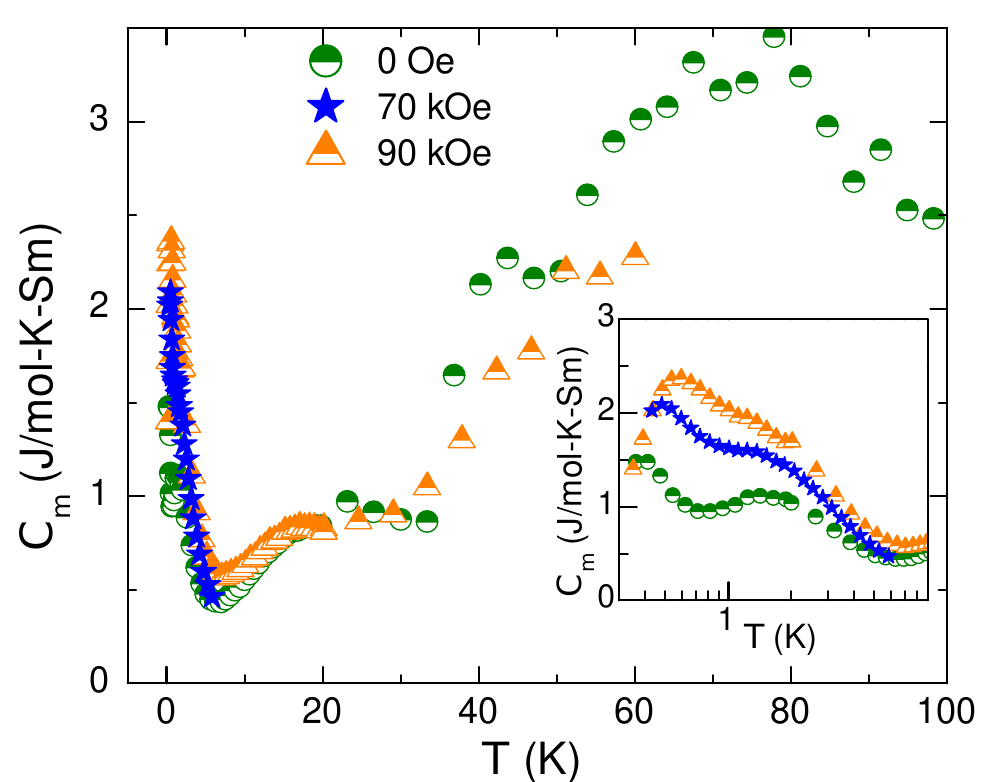}
\caption{Magnetic heat capacity $C_m(T)$ of SSZO is shown as a function of $T$ at zero field, 70\,kOe and 90\,kOe. In the inset: an enlarge view of the magnetic heat capacity $C_m(T)$ of SSZO is shown (in semi-log scale) as a function of $T$ at various applied magnetic fields.}
\label{fig:Fig-5}
\end{figure}

To extract the magnetic heat capacity $C_m$ of SSZO, we have subtracted $C_p(T)$ data of LSZO from the total $C_p(T)$ data of SSZO. We first  scaled the temperatures of the   $C_p(T)$ dataset of LSZO by multiplying the temperature by the ratio $\theta_{SSZO}/\theta_{LSZO}$ =  0.91, where $\theta_{SSZO}, \theta_{LSZO}$ are Debye temperatures of SSZO and LSZO, respectively~\cite{Bouvier_PRB_1991}. To determine  $\theta_{SSZO}$ and  $\theta_{LSZO}$, we have fitted the $C_p(T)$ data of SSZO and LSZO each with a Debye function\cite{ashcroft1976solid} in the 2-30\,K range. In Fig.\,\ref{fig:Fig-5}, $C_m(T)$ data of SSZO is shown in zero field, 70\,kOe and 90\,kOe as a function of $T$. Focusing on the zero field data, we find a few broad maxima at high temperatures ($T\geq20$\,K).  The higher temperature anomalies likely arise from excitations to the $J=7/2$ multiplet of the spin-orbit split levels. The lower temperature ($\sim$ 20 K) anomaly might be from excitations between the crystal field associated with  $\Gamma_8$ quartet and the $\Gamma_7$ doublet arising from the $J=5/2$ state.  On further lowering the temperature  below 20\,K, $C_m$  shows an upturn at $T\sim7$\,K and then a broad maximum at $T\sim1.5$\,K which is likely due to the establishment of short-range ordering (SRO) caused by interaction within the Kagom\'{e} planes. On further lowering the temperature $C_m$ shows a peak at $T\sim 0.5$\,K in zero field, which indicates establishment of LRO in SSZO likely due to inter Kagom\'{e}-plane interactions. The peak at $T\sim0.5$\,K looks reasonably sharp (though perhaps not like a classic lambda anomaly) and it surely can't be called a broad peak as is there in low dimensional systems. Our ZFC-FC $M-T$ measurements in a 50\,Oe field did not show any bifurcation down to 2\,K. However, in the absence of, say, ac $\chi$ data in this temperature range, one can't discount the possibility of a spin-glass type of transition for the 0.5\,K anomaly. On increasing the field the broad maximum gets broader and shifts to higher temperatures. Also the sharp anomaly gets sharper and shifts to higher temperatures (see Fig.\,\ref{fig:Fig-5} inset).

\begin{figure}[h!]
\centering
\includegraphics[width=9cm, height=7cm]{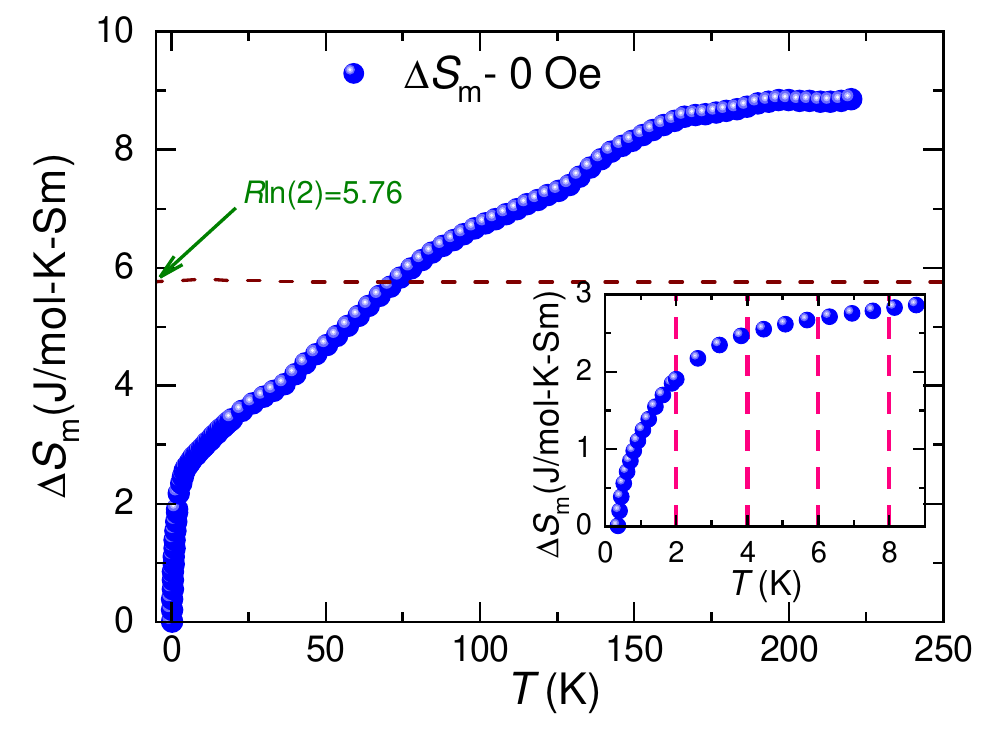}
\caption{The magnetic entropy change $\Delta S_m$ is shown as a function of $T$ for SSZO in zero field and the horizontal dashed line shows the value of $\Delta S_m$ at $R$ln(2) for $J=1/2$. The inset shows the data in the limited range up to 9 K in zero field.}
\label{fig:Fig-6}
\end{figure}

In Fig.\,\ref{fig:Fig-6} the magnetic entropy change ($\Delta S_m$) in zero field is shown as a function of $T$ for SSZO. The $\Delta S_m$ is calculated by integrating $C_m/T$ with respect to $T$. The entropy change on traversing the 0.5 K peak is around $9\%$ of $R\ln(2J+1)$ (with $J=\frac{1}{2}$).  Considering the data upto about 10 K, the entropy change is about 50\% of $R\ln2$ which is still smaller than expected. This discrepancy is probably due to the lack of data down to sufficiently lower temperatures below the ordering temperature of about 0.5 K. Indeed, if we extract the entropy change from the data in a 90 kOe field (where the maximum is shifted to a higher temperature), we get nearly 90\% of $R\ln2$. Another possibility is that of a second transition at temperatures lower than 0.35 K which further reduces the entropy{\color{red}\cite{Mauws_2018_PRB(R)}}. As SSZO has a Kagom\'{e} structure with antiferromagnetic interactions, the possibility of a glassy phase can not be ruled out given that there could be weak interplane and next near neighbour interactions which become operative at low temperatures and help relieve the frustration to some extent and result in a glassy state. This also could account for the residual entropy. The plateaus seen at  temperatures higher than 20 K likely arise from transitions to the higher multiplets.

\section{Conclusion}

In summary, we have prepared a single phase polycrystalline sample of Sm$_3$Sb$_3$Zn$_2$O$_{14}$ which has magnetic Kagom\'{e} planes formed by Sm$^{3+}$ ions. From the Curie constant determined from the low-temperature susceptibility data, we find an effective moment consistent with a pseudo $S=1/2$ state. The small value of $\theta_{CW}$ (-1.4\,K) from the low-temperature Curie-Weiss fit suggests weak antiferromagnetic interactions.  Specific heat data (which were taken down to 0.35 K; lower than the 1.8 K limit of $\chi(T)$ data) show a sharp anomaly at about 0.5 K  which indicates LRO (or possibly a glassy state). Low temperature ac susceptibility measurements might help resolve this. It would also be useful to do local probe measurements like muon spin relaxation to validate this conclusion and to further check for the prevalence of fluctuations in the low-temperature regime.

%%---------------------------------------------
%                     Acknowledgments
%%---------------------------------------------
\section{Acknowledgment}
Vinod Kumar would like to acknowledge the financial support and central facility measurements for MPMS and PPMS provided by IITB and IRCC. Vinod Kumar would also acknowledge the fruitful discussion with Mr. Swayam Kesari, scientist at BARC, Mumbai, Maharashtra, India.

%%---------------------------------------------
%                     Reference Page
%%---------------------------------------------

\bibliographystyle{apsrev4-1}
\bibliography{bib}

\end{document}